\documentclass[12pt]{article}
\usepackage{color}
\usepackage{amsmath,amssymb,amsfonts,dcolumn,color,graphicx,graphics,latexsym,epsfig}

\def\beq{\begin{equation}}
\def\eeq{\end{equation}}
\def\barr{\begin{array}}
\def\earr{\end{array}}
\def\dis{\displaystyle}
\begin{document}
\thispagestyle{empty}

\begin{center}

{\Large\bf Black hole in a radially slant warped braneworld scenario}\\ [25mm]

Sayantani Lahiri\footnote{E-mail: sayantani.lahiri@gmail.com}\\
{\em Relativity and Cosmology Centre 
\\ Department of Physics}
\\{\em Jadavpur University
\\ Raja Subodh Chandra Mullick Road
\\Jadavpur
\\Kolkata- 700032, India}  \\ [15mm]

\end{center}

\bigskip

\abstract{As a follow-up to our previous paper arxiv: 1309.4244[hep-th], 
we determine radion induced spherically symmetric solution using the gradient approximation scheme,
when two warped $3$-branes are slant with respect to each other 
such that the radion field in this case is a radial co-ordinate varying function. The slanting between the branes is assumed to be small.
In the absence of any on-brane matter except that induced effects of radion field, the resulting black hole is found to be extremal Reissner-Nordstrom whose mass is proportional to the slanting between the $3$-branes. It is however seen that the 
inter-brane separation diminishes to zero for a particular value of radial co-ordinate which can be mimicked as intersection of the branes. This situation can be circumvented by adding traceless on-brane matter on the negative tension brane.}

\newpage

\section{Introduction}
It has been over a decade now that braneworld models have been introduced 
in which our Universe and all Standard model fields are 
assumed to be located on an hypersurface which is a $3$-brane embedded in an higher dimensional bulk. 
In these braneworld models, 
our four dimensional world is described by low energy effective theory 
so that the TeV scale physics are reproduced on the 
$3$-brane which also carries the projected 
information of the bulk in the form of effective energy momentum tensor on the brane.\\
We, in this paper, solely concentrate on Randall-Sundrum (RS) \cite{randall} braneworld model 
consisting of two $3$-branes possessing equal and opposite brane tensions, 
primarily proposed to resolve 
the naturalness problem or gauge hierarchy problem in particle physics. 
The RS model was immensely successful in 
resolving the hierarchy problem without introducing any intermediate scale in the theory. 
But due to the presence of equal and opposite brane tensions, 
the inter-brane separation distance 
was susceptible to instability which was stabilized by introducing a scalar field in the bulk. 
This technique, in addition, does not introduce any new fine-tuning in the model. 
Moreover. various implications of RS model have been 
extensively studied in the areas of 
cosmology and astrophysics and particle phenomenology \cite{maartens,phen,lang}. \\ 
The present paper has been developed in the 
track of \cite{rad_ind} where we assume that the two branes in the 
Randall Sundrum model are not parallel instead 
they are slant by an angle $\alpha$ with respect to each other such that 
the slanting between the branes is measured by a factor $\alpha f(\rho)$, where $\rho$ is an
isotropic co-ordinate. 
Here $\alpha$ being a proportionality constant, 
is alternatively called the slant parameter and $f(\rho)$ may be called the slant function. 
With the given set-up of the model, the radion measuring 
the inter-brane separation emerges to be a radial co-ordinate 
dependent field to an observer residing on either of the two $3$-branes. 
As a result, 
even in the absence of matter densities and pressures on the branes, 
the radion field acts as a source of effective energy momentum tensor, 
thereby making the branes non-flat.\\
Starting with a brief introduction, 
in Section 2 we have briefly discussed the gradiant expansion scheme 
that gives the formulation for determining the low energy effective Einstein's equations on 
non-flat branes in the presence of a spacetime dependent radion field.\\
In Section 3 we have shown 
that an extremal Reissner-Nordstrom black hole solution is generated 
on the visible brane under the influence of radial co-ordinate varying modulus field and
in the absence of matter on either of the branes. 
The existence of such a black hole solution has been shown to be 
purely a consequence of slanting between the two $3$- branes. 
The entire study is carried out 
with the help of low energy effective theory approach 
based on a perturbative scheme developed by Kanno and Soda \cite{kanno}. 
It is found that the black mass can be explicitly carries an $\alpha$ dependence 
with a proper scaling. 
After obtaining the solution of the radion field, 
we derived the slant function $f(\rho)$ in this section. \\
However, 
it is found that the radion field is plagued with an 
instability problem which can be eliminated by 
introducing brane matter on either one of the two branes. \\
Finally we write the concluding remarks of our work in section 4.

\section{Low energy effective theory : gradiant approximation scheme}
Let us briefly describe gradiant approximation scheme which gives the low energy effective theory of a two brane system \cite{kanno}.
Let us assume the following five dimensional action :\\[1mm]
\underline{\textbf{Action}} :\\
\beq
S\,=\,\frac{1}{2 \kappa^2}\int d^5 x \sqrt{-g}\left( \textit{R} +\dis\frac{12}{l^2}\right)-\dis\Sigma_{i = a,b}\,\sigma_{\textit{i}} \int d^4 x \sqrt{-g^{\textit{i}\,brane}}\,+\,\dis\Sigma_{i = a,b} \int d^4 x \sqrt{-g^{\textit{i}\,brane}}\,L^{\textit{i}}\,_{matter}
\eeq
where $\kappa^2$ is the $5$D gravitational coupling constant and  $l$ is the bulk curvature radius. The two $Z_2$ symmetric $3$- branes placed at orbifold fixed points i,e. $y=0$ and $y=l$ are embedded in five dimensional AdS bulk. The positive tension $3$-brane placed at $y=0$ is known as the Planck brane ($i=a$) while the $3$-brane possessing negative brane tension known as the visible brane ($i=b$), is located at $y=l$ with energy scales typically around TeV . The geometrical structure of the spacetime is described by $ M^{(1,3)} \times S_1/Z_2$.\\
The most general five dimensional metric for two non-flat branes with a spacetime dependent inter-brane separation distance is given by :\\[1mm]
\beq
ds^2\,=\,e^{2 \tilde{\phi}(x)}\, dy^2 \,+\,g_{\mu \nu}(y, x^{\mu})\,dx^{\mu} dx^{\nu}
\eeq
The astrophysical solutions which is our present area of study occur at the length scales much large compared to the bulk curvature and hence can be investigated in the regime of low energy effective theory. 
The non-zero contribution of the radion field on a given $3$-brane appears through effective Einstein's equations.
To determine the effective Einstein's equations, at first one needs to solve the bulk Einstein's equations.
However, it is extremely difficult to solve the five dimensional bulk Einstein's equations exactly in the presence of the radion field without invoking to some perturbative approach. 
In the low energy regime the brane curvature radius $L$ is large compared to bulk curvature $l$. By taking into consideration of this fact, a peturbative approach has been adopted by Kanno and Soda, where both the bulk metric $g_{\mu \nu}$ and the extrinsic curvature $K^{\mu}_{\nu}$ are expanded in a perturbative series with an increasing order of dimensionless perturbation parameter $\epsilon$ such that $\epsilon = (\frac{l}{L})^2 << 1$. 
In this metric based iterative process, the effective Einstein's equations on a given $3$-brane are obtained at the first order of perturbation theory.
At the zeroth order the brane curvature and on brane matter are neglected. As a result, each of the branes are characterised by intrinsic brane tensions. The junction conditions at this order give rise to a condition between the bulk curvature radius and the brane tensions exactly identical to Randall Sundrum fine-tuning condition. The two $3$-branes are therefore flat and static separated by a constant radion field. \\ [1mm]
At first order, the curvature on the branes are taken into consideration and the effect of the radion field comes into the picture. With the solutions of extrinsic curvature, induced metric  of the zeroth order and junction conditions yield the effective Einstein's equations on the visible brane as follows  :
\beq
^{(\,4\,)}G^{\mu}_{ \nu}\,=\,\frac{\kappa^2}{l \Phi} \, T^{b\,\mu}\,_{\nu}\,+\,\frac{\kappa^2(1\,+\,\Phi)}{l \Phi}\,T^{a\,\mu}\,_{\nu}\,+\,\frac{1}{\Phi}\left(\widetilde{\nabla}^{\mu}\widetilde{\nabla}_{\nu}\Phi\,-\,\delta^{\mu}_{\nu}\,\widetilde{\nabla}^{\alpha}\,\widetilde{\nabla}_{\alpha}\Phi\right)\,+\,\frac{\omega(\Phi)}{\Phi^2}\,\left(\widetilde{\nabla}^{\mu}\Phi\,\widetilde{\nabla}_{\nu}\Phi\,-\,\frac{1}{2}\delta^{\mu}_{\nu}\,\widetilde{\nabla}^{\alpha}\Phi\,\widetilde{\nabla}_{\alpha}\Phi\right)        \label{EE_eff}
\eeq
and the corresponding equation of motion of the scalar field on the negative tension brane is given by,
\begin{equation}
\widetilde{\nabla}^{\alpha}\widetilde{\nabla}_{\alpha}\Phi =\frac{\kappa^2}{l}\frac{T^a + T^b}{2\omega+3} - 
\frac{1}{2\omega +3} \frac{d\omega}{d\Phi} (\widetilde{\nabla}^{\alpha}\Phi)(
\widetilde{\nabla}_{\alpha}\Phi )   \label{scl_eqn}
\end{equation}
Here $T^{a}$ and $T^{b}$ are traces of energy momentum tensors on respective Planck brane and visible brane.
Here, $\Phi =e^{2 d/l}-1$ and  $\widetilde{\nabla}_{\nu}$ is the covariant derivative with respect to the induced metric of the visible brane $f_{\mu \nu} $. The spacetime varying proper distance $d$ between the two $3$-branes is defined as :
\beq
d(x) = \int _{0} ^{l} e^{\tilde{\phi}(x)}\, dy
\eeq
The Einstein's equations on the negative tension brane suggest a quasi-scalar tensor theory with the coupling function $\omega(\Phi)\,=\,-\dis\frac{3}{2}\,\dis\frac{\Phi}{1 + \Phi} $.\\

\section{Spherically symmetric solution in slant braneworld}

The well-known metric ansatz for two parallel $3$-branes separated by a constant brane separation distance $R$ is given by :
\beq 
ds^2\,=\,dy^2 + a^2(x^{\sigma},y)g_{\mu \nu}(y, x^{\sigma})dx^{\mu} dx^{\nu}    \label{5D_metric}
\eeq 
Here $y = R \phi$ and $\phi$ runs from $0$ to $\pi$ due to orbifold compactification.
In this work, we choose isotropic co-ordinate system $(t, \rho, \theta, \phi)$ to 
express the $5$D metric as well as $4$D induced metric. 
It is to be noted that with a specific co-ordinate transformation we can always switch 
to $(t, r, \theta, \phi)$ co-ordinate system, where $r$ is the radial co-ordinate. 
This transformation in our case will be discussed subsequently.  \\
Let us consider that two $3$-branes, embedded in five dimensional AdS bulk, are slightly slant with respect to each other such that the slanting is proportional to $f(\rho)$, where $f(\rho)$ may be called the slant function. This implies that the inter-brane separation distance does not remain constant instead it becomes a $\rho$ varying function. In this set-up, the five dimensional metric may be taken to be :\\[1mm]
\textbf{\underline{$5$D Metric Ansatz for slant braneworld set-up}} :
\beq
ds^2\,=\,\left(1 + \dis\frac{\alpha f(\rho)}{R^2}\right)\, dy^2 \,+a^2(\rho,y)\,g_{\mu \nu}(y, x^{\mu})\,dx^{\mu} dx^{\nu}     \label{5D_metric_1}
\eeq
where the slant parameter $\alpha$ which measures the degree of slanting between the two branes is assumed to be small and $a(\rho,y)$ is the warp factor. When the branes are parallel i,e. $ \alpha = 0$, the inter-brane distance  becomes equal to constant value $R$ and the above metic choice reduces to eqn.{(\ref{5D_metric})}. But in slant braneworld scenario, the proper distance within the interval $y = 0$ to $y = l$ is a co-ordinate varying function given by :
\beq
d(\rho)\,=\,\int_{0}^{l} \sqrt{1 + \dis\frac{\alpha f(\rho)}{R^2}}\, dy \,=\,\sqrt{1 + \dis\frac{\alpha f(\rho)}{R^2}}\,l  \label{prop_len}
\eeq
The warp factor is expressed in terms of isotropic co-ordinate $\rho$, such that,
\beq
a(y,\rho)\,=\, e^{\left[-\sqrt{1 + \dis\frac{\alpha f(\rho)}{R^2}}\dis \frac{y}{l}\right]}\,=\,e^{-\dis \frac{d(\rho)}{l}y}
\eeq
Now in order to generate a black hole solution on the visible brane, we take following metric  which is identical to Majumdar-papapetrou choice of metric.\\ [1mm]
\underline{\textbf{$4$D metric choice }} :\\[1mm]
\beq
ds^2_{(4)}\,=\,-\frac{1}{U^2(\rho)}dt^2 + U^2(\rho)[d\rho^2\,+\,\rho^2 d\Omega^2]       \label{BH1}
\eeq
The radion field which is a scalar field on the visible brane is given by 
\beq
\Phi(\rho) = e^{2d(\rho)/l} - 1
\eeq
where $l$ is the bulk AdS curvature and $d(\rho)$ is the proper length between the two $3$-branes as given by eqn.{(\ref{prop_len})}.  
Since our aim is to generate an astrophysical black hole solution on a $3$-brane with the radion field as the only source of effective energy momentum tensor so we initially neglect the brane matter $T^{a\,\mu}\,_{\nu}$ and $T^{b\,\mu}\,_{\nu}$ in eqn.{(\ref{EE_eff})}. By putting $T^{a\,\mu}\,_{\nu} = 0 = T^{b\,\mu}\,_{\nu}$ and using eqn.{(\ref{BH1})}, the equation of motion of the scalar field eqn.{(\ref{scl_eqn})} on the visible brane is given by,
\beq
\dis \frac{2 f'}{r} + f'' \,=\, \dis \frac{\alpha^2 f'^2}{2 R^2}\left[\dis \frac{1}{\sqrt{1 + \dis \frac{\alpha f}{R^2}}} - \dis \frac{1}{\left(1 + \dis \frac{\alpha f}{R^2}\right)} \right]
\label{sf_eq}
\eeq
where $'$ is the derivative with respect to $\rho$. Now on integrating the above equation, we get,
\beq
\dis \frac{\exp{\left[\sqrt{1 + \dis \frac{\alpha f}{R^2}}\right]}}{\sqrt{1 + \dis \frac{\alpha f}{R^2}}}\,f' \,=\,\dis \frac{2 \tilde{C_3}}{\rho^2}   \label{eqn_2}
\eeq
In the absence of brane matter, we obtain following Einstein's equations by substituting eqn.{(\ref{BH1})} in eqn.{(\ref{EE_eff})} : \\[2mm]
\underline{tt component :}\\[1mm]

\beq
\dis\frac{4}{\rho}\frac{U'}{U}-\left(\dis\frac{U'}{U}\right)^2 + 2\frac{U''}{U} \, = \,- A_1(\rho) \left[\dis \frac{\alpha}{R^2 \sqrt{1 + \dis\frac{\alpha f}{R^2}}} \dis \frac{f'}{2} \dis \frac{U'}{U} - \frac{\alpha^2}{R^4 \left(1 + \dis \frac{\alpha f}{R^2} \right)} \dis \frac{f'^2}{8}\right]  \label{tt_comp}
\eeq

\underline{$\rho \rho$ component :}\\[1mm]

\beq
-\left(\dis\frac{U'}{U}\right)^2 = -A_1(\rho)\left[\dis \frac{\alpha}{R^2 \sqrt{1 + \dis \frac{\alpha f}{R^2}}} \left(\dis \frac {f'}{\rho} + \dis \frac{f'}{2} \dis \frac{U'}{U}\right) + \dis \frac{ \alpha^2}{R^4 \left(1 + \dis \frac{\alpha f}{R^2} \right)} \dis \frac{3 f'^2}{8}\right]         \label{rho_comp}
\eeq

\underline{$\theta\theta$ and $\tilde{\phi}\tilde{\phi}$ components :}
\beq
\left(\dis\frac{U'}{U}\right)^2 = - A_1 (\rho) \left[-\dis \frac{\alpha}{R^2 \sqrt{1 + \dis \frac{\alpha f}{R^2}}} \left(\dis \frac {f'}{2 \rho} + \dis \frac{f'}{2} \dis \frac{U'}{U}\right) - \dis \frac{ \alpha^2}{R^4 \left(1 + \dis \frac{\alpha f}{R^2} \right)} \dis \frac{f'^2}{8} \right]              \label{theta_comp}
\eeq
where $A_1(\rho)\,=\,  \exp{\left[2 ( 1 + \dis\frac{\alpha f(\rho)}{R^2})^{1/2}\right]} \left[-1 + \coth \sqrt{1 + \dis\frac{\alpha f(\rho)}{R^2}}\right]$.\\
Addition of eqn.{(\ref{rho_comp})} and eqn.{(\ref{theta_comp})} results into,
\beq
\dis \frac{\alpha}{R^2 \sqrt{1 + \dis \frac{\alpha f}{R^2}}} \dis \frac{f'^2}{2} + \dis \frac{f'}{\rho}\,=\,0   \label{eqn_1}
\eeq
which when substituted in the Einstein equations [eqn.{(\ref{tt_comp})} - eqn.{(\ref{theta_comp})}] give rise to only two independent equations :  \\[1mm]
\underline{tt component :}\\[1mm]
\beq
\dis\frac{4}{\rho}\frac{U'}{U}-\left(\dis\frac{U'}{U}\right)^2 + 2\frac{U''}{U} \, = \,\dis \frac{A_1(\rho)}{\sqrt{1 + \dis \frac{\alpha f}{R^2}}} \dis \frac{\alpha}{R^2}\left[-\dis \frac{f'}{2} \frac{U'}{U} - \frac{f'}{4 \,\rho}\right]                    \label{tt_1}
\eeq

\underline{$\rho \rho$ component :}\\[1mm]
\beq
\left(\dis\frac{U'}{U}\right)^2 = \dis \frac{A_1 (\rho)}{\sqrt{1 + \dis \frac{\alpha f}{R^2}}}\dis \frac{\alpha}{R^2} \left[\dis \frac{f'}{2} \frac{U'}{U} + \frac{f'}{4 \,\rho} \right]   \label{rho_1}
\eeq
where $\theta \theta$ and $\tilde{\phi}\tilde{\phi}$ components are identical to $rr$ component. Finally adding eqn.{(\ref{tt_1})} and eqn.{(\ref{rho_1})} we obtain the following second order differential equation of $U(\rho)$ as,
\beq
U'' + 2 \dis \frac{U'}{\rho} \,=\, 0   \label{eq_5}
\eeq
which is the Laplace's equation $ \widetilde{\nabla}^2 U(\rho) = 0 $. So the solution of black hole metric $U(\rho)$ is given by :
\beq
U(r)\,=\, C_1 + \dis \frac{C_2}{\rho}     \label{eqn_3}
\eeq 
where $C_1$ and $C_2$ are integration constants. Being a dimensionless quantity, we put $C_1 = 1$.
Let us now try to determine the mass of the obtained black hole solution which involves the integration constant $C_2$. For this, we need to first find the solution of the scalar field $\Phi(\rho)$. Since $f'(\rho) \neq 0$, therefore eqn.{(\ref{eqn_1})} and eqn.{(\ref{eqn_2})} give  $\Phi(\rho)$ as :
\beq
\Phi(\rho)\,=\,\exp{{\left[2\,\sqrt{1 + \dis \frac{\alpha f(\rho)}{R^2}}\right]}} -1 \,=\,\dis \frac{C_3 ^2}{\rho^2} -1   \label{radion}
\eeq
where $C_3^2 = \dis \frac{\tilde{C_3}^2 \alpha^2}{R^4}$ and $C_3$ has dimension of $[L]$. With the solutions of $\Phi(\rho)$, $U(\rho)$ and their derivatives, when substituted in either eqn.{(\ref{tt_1})} or eqn.{(\ref{rho_1})} we obtain a condition between $C_2$, $C_3$ and $\alpha$ as given below,
\beq
C_3 \,=\, C_2\,=\,\dis \frac{\tilde{C_3} \alpha}{R^2} 
\eeq
Since $\tilde{C_3}$ has dimension of $[L]^2$, so by scaling $\tilde{C_3} = \beta R^2$, where $\beta$ is a dimensionless proportionality constant, let $C_3 = \beta \alpha = \gamma$, so that the solution of the black hole metric in isotropic co-ordinates is given by :
\beq
ds^2_{(4)}\,=\,-\dis \frac{1}{\left(1 + \dis \frac{\gamma}{\rho}\right)^2} \,dt^2 + \left(1 + \dis \frac{\gamma}{\rho}\right)^2 [d\rho^2 + \rho^2 d\Omega^2 ] \label{eqn_6}
\eeq
Thus $U(\rho)$ satisfying the Laplace's equation ( given by eqn.{(\ref{eq_5})}) gives rise to an extremal Reissner-Nordstrom (RN) solution where the charge and the mass of the black hole are identical. 
Furthermore, the solution suggests that in the absence of brane matter on either of the two $3$-branes, 
the radion field is solely responsible for generating the mass of the black hole. 
Here, the slant parameter $\alpha$ disguised as $\gamma$ plays the role of mass of the black hole where we have $U(\rho) = 1 + \dis\frac{\gamma}{\rho}$.
It is to be noted that the effective energy momentum tensor contributed by the slanting between the branes is itself traceless.\\ 
Now, from the solution of radion field, we can readily determine the slant function $f(\rho)$ as:
\beq 
f(\rho)\,=\,\dis \frac{R^2}{\alpha}\left[\left( \ln \dis \frac{\gamma }{\rho}\right)^2 -1 \right]
\eeq 
So, with the solution of the slant function $f(\rho)$, the proper distance becomes :
\beq
d(\rho) =  l\, \ln \dis\frac{\gamma}{\rho}
\eeq
Similarly, with $C_3 = \gamma$, the radion field in terms of slant parameter $\gamma$ can be written as :
\beq
\Phi(\rho) = \dis\frac{\gamma^2}{\rho^2} - 1
\eeq
 However, it is always possible to define a co-ordinate transformation such that : $r = \rho + \gamma$ which relates the radial co-ordinate $r$ and isotropic co-ordinate $\rho$. Using this transformation relation, the metric solution for the extremal black hole can be re-expressed as :
\beq
ds^2 \,=\,-\left(1 - \frac{\gamma}{r}\right)^2 dt^2 + \dis\frac{1}{\left(1 - \dis\frac{\gamma}{r}\right)^2}\,d r^2 + r^2 d\Omega^2 
\eeq
where $\gamma$ proportional to slant parameter $\alpha$ plays the mass of the extremal RN black hole.\\[2mm]
By assuming that the slant parameter $\alpha$ is a positive quantity, 
the horizon is found to be located at :
\beq
r_{H} = \gamma
\eeq
which corresponds to $\rho =0$ where the inter-brane separation distance $d(\rho)$ is undefined. 
This scenario can be visualized in the following way:\\
As $\rho$ increases from zero, the proper distance $d(\rho)$ 
between the two branes remains positive but decreases logarithmically, 
until becomes zero at $\rho=\gamma$, where the $3$-branes intersect. 
For $\rho > \gamma$, the effective distance between the intersecting branes 
become negative as it is now measured in the opposite direction
due to change of location of the branes beyond the intersection point.
Once again, when $\rho$ exceeds the value $\gamma$, the proper distance becomes negative in other words, the two branes re-emerge after intersection. \\
So, the proper distance $d(\rho)$ remains positive for $0<\rho<\gamma$ which corresponds to $\gamma< r < 2 \gamma$. In terms of radial co-ordinate $r$, the proper distance is:
\beq
d(r)=l \ln \dis\frac{\gamma}{r-\gamma}
\eeq
Here we note that in the limit $\gamma \rightarrow 0$ 
i,e. when the inter-brane angle vanishes, the two $3$-branes become once again parallel 
and flat Minkowski metric is recovered. \\[1mm]
Before concluding our work, 
we briefly address the stability aspect of the model where we find that
the radion field $\Phi$ from eqn.{(\ref{radion})} and 
 the proper distance $d$ vanishes an case  $\dis \frac{C_3}{\rho} = 1$ i,e. when $\rho = \gamma$. 
Following our previous work regarding this issue, 
the possible brane intersection can be circumvented by 
introducing traceless matter 
only on the visible brane 
(Planck brane matter : $T^{a} \,^{\mu}\, _{\nu} = 0$) to keep the proper distance finite 
and non-zero. 
The brane matter in this case which obeys following relation :
\beq
-\rho_b + \tau_b + 2 p_b \,=\,0
\eeq
The tracelessness property of the brane matter keeps 
the equation of motion of the scalar field given by eqn.{(\ref{sf_eq})} 
unchanged and therefore eqn.{(\ref{eqn_2})} 
can be expressed in terms of first derivative of $\Phi(\rho)$ as :
\beq
\dis \frac{\Phi'(\rho)}{\sqrt{1 + \Phi(\rho)}}\,=\, \dis \frac{2 C_3}{\rho}
\eeq
where $C_3 =\gamma$. The most general solution of the radion field as a result is :
\beq
\Phi(\rho)\,=\, \left(\dis\frac{\gamma}{\rho} + \dis \frac{C_4}{2}\right)^2 -1
\eeq
where $C_4$ is the integration constant.The stabilization of  the inter-brane separation distance 
is achieved if $C_4 > 2$ is taken. 
It is to be further noted that using Einstein's equations, the matter density 
and pressure can be evaluated 
in terms of constants $\gamma$, $C_4$. \\
 When $C_4 =0$, both the branes are empty and
the solution of the radion field given by eqn.{(\ref{radion})}. 
Moreover, in the presence of traceless on-brane matter and  non-zero slant parameter, the black hole metric $U(\rho)$ still obeys the extremal RN solution which can be easily checked from the non-vacuum effective Einstein's equations given by eqn.{(\ref{EE_eff})}.
However, mass of the extremal black hole in this case 
will be redefined 
due to existence of both traceless on-brane matter and as well as non-zero 
angle between the two $3$-branes. \\

\section{Conclusion}

The radion field induced spherically symmetric solution in a radially 
slant braneworld model is found to be necessarily extremal Reissner-Nordstrom on the
visible brane. The black hole solution is obtained by considering Majumdar-Papapetrou choice of metric in isotropic co-ordinate system. The solution of slant function is determined in this case. The proper distance and the radion field are found to be proportional to the slant parameter.\\
It is seen that the co-ordinate transformation which relates $\rho$ and $r$ is proportional to the slant parameter $\gamma$. The black hole mass and the event horizon are both found to be related to the slant parameter. It is seen that the proper distance $d(\rho)$ after remaining
positive and non-zero, vanishes for a particular value when $\rho=\gamma $, where $\gamma$ is the slant parameter scaled by a dimensionless proportionality constant. 
It has been shown that the collision of the two $3$-branes can be avoided by introducing traceless matter on the visible brane. It may be noted that the static limit of the model corresponds to $\gamma \rightarrow 0$ when Minkowski metric is recovered.   

\section*{Acknowlegment}
I sincerely thank Soumitra SenGupta for suggesting this problem to me. 
The discussions I had with him have been extremely
helpful for me for completion of this work.


\begin{thebibliography}{99}

\bibitem{randall} L. Randall and R. Sundrum, 
Phys. Rev. Lett.  {\bf 83} 3370 (1999); 
{\it ibid} {\bf 83} 4690 (1999)


\bibitem{maartens} R. Maartens and K. Koyama, Living Rev. Relativity {\bf 13},5 (2010).


\bibitem{phen} A. J. Buras, B. Duling and S. Gori, JHEP 0909:076,  (2009);
A.V. Kisselev JHEP 0809:039 (2008); JHEP 0703:006, (2007); 
M. C. Kumar, P. Mathews, V. Ravindran Eur.Phys.J.C49, (2007);
P. Osland, A. A. Pankov, A. V. Tsytrinov and N. Paver, AIP Conf.Proc.1317:201-208,  (2011); 
K. Agashe, A. Belyaev, T. Krupovnickas, G. Perez and J. Virzi, Phys. Rev. D {\bf77} 015003, (2008); M. J. Duff and J. T. Liu , Phys. Rev. Lett. 85 2052 (2000); 
H. Boschi-Filho and N. R. F. Braga, Nucl. Phys. B {\bf608}:319-332,(2001)



\bibitem{lang}            P. Bin´etruy, C. Deffayet, and D. Langlois, Nucl. Phys. B {\bf 565}, 269 (2000);
P. Bin´etruy, C. Deffayet, U. Ellwanger, and D. Langlois, Phys. Lett. B {\bf 477}, 285 (2000);
N.Kaloper  Phys. Rev  D {\bf 60} ,123506 (1999);  I. P. Neupane  Phys. Lett. B {\bf 683}, (2010); 
T. Nihei, Phys. Lett. B {\bf 465}, 81 (1999); 
J.M. Cline, C.Grojean, and G. Servant, (1999), Phys. Rev. Lett. {\bf 83};
C. Csaki, M. Graesser, C. Kolda and J.Terning, (1999). Phys. Lett. B {\bf 462};
T. Multamaki, I. Vilja ;   Phys. Lett. B {\bf 559} (2003);
Sahni, Y. Shtanov ;   JCAP 0311 (2003) 014;  
K. Koyama, Gen. Rel. Grav.{\bf 40}:421-450, 2008.


\bibitem{rad_ind}  Sayan Kar, Sayantani Lahiri, Soumitra SenGupta;
Phys. Rev. D{\bf 88} 123509 (2013)

\bibitem{kanno} S. Kanno, J. Soda, Phys.Rev. {\bf D 66}, 083506 (2002)



\end{thebibliography}
\end{document}